\documentclass[preprintnumbers,amsmath,amssymb,showpacs]{revtex4}

\usepackage{graphicx}
\usepackage{dcolumn}
\usepackage{bm}

\begin{document}

\title{Tighter monogamy relations of multiqubit entanglement in terms of R\'{e}nyi-$\alpha$ entanglement}

\author{Limin Gao}
\author{Fengli Yan}
\email{flyan@hebtu.edu.cn}
\affiliation {College of Physics Science and Information Engineering, Hebei
Normal University, Shijiazhuang 050024, China}
\author{Ting Gao}
\email{gaoting@hebtu.edu.cn}
\affiliation {College of Mathematics and Information Science, Hebei
Normal University, Shijiazhuang 050024, China}

\date{\today}

\begin{abstract}
We present a class of tight monogamy relations in terms of R\'{e}nyi-$\alpha$ entanglement, which are tighter than the monogamy relations of multiqubit entanglement just based on the power of the R\'{e}nyi-$\alpha$ entanglement for $\alpha\geq2$ and the power $\eta>1$. For $2>\alpha\geq\frac{\sqrt{7}-1}{2}$ and the power $\eta>2$, we establish a class of tight monogamy relations of multiqubit entanglement with larger lower bounds than the existing monogamy relations of multiqubit entanglement.
\end{abstract}


\maketitle

\section{INTRODUCTION}

A key property of entanglement is known as monogamy relations [1, 2], that is, entanglement cannot be freely shared unconditionally among the multipartite quantum systems. The first mathematical characterization of monogamy relation was known as the monogamy inequality for three-qubit quantum state in terms of squared concurrence, and it is called CKW-inequality [1]. Furthermore, Osborne and Verstraete generalized this monogamy inequality to arbitrary multiqubit systems [3]. Later, the same monogamy inequality was also generalized to other entanglement measures such as entanglement negativity [4] and entanglement of formation [5]. Monogamy relations are used to characterize the distribution of entanglement in multipartite systems. Moreover, the monogamy property has many important applications in quantum information theory [6], condensed-matter physics [7] and even black-hole physics [8].

As a generalization of entanglement of formation, the R\'{e}nyi-$\alpha$ entanglement is a well-defined entanglement measure, and has been widely used in the study of quantum information theory [9, 10]. It has been shown that if $\alpha\geq2$, the R\'{e}nyi-$\alpha$ entanglement satisfies the monogamy relations in $N$-qubit systems [11]. When $\alpha\geq\frac{\sqrt{7}-1}{2}$, the squared R\'{e}nyi-$\alpha$ entanglement satisfies the monogamy relations in $N$-qubit systems [12]. Recently, a class of tight monogamy relations were derived in multiqubit systems [13-16]. In this paper, we establish a class of tight monogamy relations of multiqubit entanglement in terms of R\'{e}nyi-$\alpha$ entanglement related to the power of the R\'{e}nyi-$\alpha$ entanglement, which are tighter than the results in [12, 15, 16].

\section{THE R\'{E}NYI-$\alpha$ ENTANGLEMENT}

The R\'{e}nyi-$\alpha$ entanglement of a bipartite pure state $|\psi\rangle_{AB}$, is defined as [17]
\begin{equation}\label{}
E_{\alpha}(|\psi\rangle_{AB})=\frac{1}{1-\alpha}\log_{2}(\text{tr}\rho_{A}^{\alpha})
\end{equation}
for any $\alpha>0$ and $\alpha\neq1$, $\rho_{A}=\text{tr}_{B}(|\psi\rangle_{AB}\langle\psi|)$. If $\alpha$ tends to 1, the R\'{e}nyi-$\alpha$ entanglement converges to the von Neumann entropy. For a bipartite mixed state $\rho_{AB}$, the R\'{e}nyi-$\alpha$ entanglement is defined via the convex-roof extension
\begin{equation}\label{}
E_{\alpha}(\rho_{AB})=\min\sum_{i}p_{i}E_{\alpha}(|\psi_{i}\rangle_{AB}),
\end{equation}
where the minimum taken over all possible pure-state decompositions of $\rho_{AB}=\sum_{i}p_{i}|\psi_{i}\rangle_{AB}\langle\psi_{i}|$.

Let us recall the definition of concurrence. For a bipartite pure state $|\phi\rangle_{AB}$, the concurrence is [18]
\begin{equation}\label{}
C(|\phi\rangle_{AB})=\sqrt{2(1-\text{tr}\rho_{A}^{2})},
\end{equation}
where $\rho_{A}=\text{tr}_{B}(|\phi\rangle_{AB}\langle\phi|)$. For a mixed state $\rho_{AB}$, the concurrence is defined via the convex-roof extension
\begin{equation}\label{}
C(\rho_{AB})=\min\sum_{j}p_{j}C(|\phi_{j}\rangle_{AB}),
\end{equation}
where the minimum taken over all possible pure-state decompositions of $\rho_{AB}=\sum_{j}p_{j}|\phi_{j}\rangle_{AB}\langle\phi_{j}|$.

For an arbitrary $N$-qubit state $\rho_{A|B_{1}\cdots B_{N-1}}\in \mathcal{H}_{A}\otimes \mathcal{H}_{B_1}\otimes\cdots\otimes\ \mathcal{H}_{B_{N-1}}$, the concurrence $C(\rho_{A|B_{1}\cdots B_{N-1}})$ of the state $\rho_{A|B_{1}\cdots B_{N-1}}$ in the partition $A$ and $B_{1}\cdots\ B_{N-1}$, satisfies [3]
\begin{equation}\label{}
C^{2}(\rho_{A|B_{1}\cdots B_{N-1}})-C^{2}(\rho_{A|B_{1}})-\cdots-C^{2}(\rho_{A|B_{N-1}})\geq0,
\end{equation}
where $\rho_{A|B_{i}}=\text{tr}_{B_{1}\cdots B_{i-1}B_{i+1}\cdots B_{N-1}}(\rho_{A|B_{1}\cdots B_{N-1}})$, $\mathcal{H}_{A}, \mathcal{H}_{B_1},\cdots, \mathcal{H}_{B_{N-1}}$ are Hilbert spaces of the systems $A, B_{1}, \cdots, B_{N-1}$, respectively.

It has been proved that [11, 19], when $\alpha\geq\frac{\sqrt{7}-1}{2}$, for a two-qubit state, the R\'{e}nyi-$\alpha$ entanglement has an analytical formula
\begin{equation}\label{}
E_{\alpha}(\rho_{AB})=g_{\alpha}(C(\rho_{AB})).
\end{equation}
Here the function $g_{\alpha}(x)$ is a monotonically increasing and convex function expressed as
\begin{equation}\label{}
g_{\alpha}(x)=\frac{1}{1-\alpha}\log_{2}[(\frac{1-\sqrt{1-x^{2}}}{2})^{\alpha}+(\frac{1+\sqrt{1-x^{2}}}{2})^{\alpha}]
\end{equation}
in $0\leq x\leq1$.

The function $g_{\alpha}(x)$ in Eq. (7) for $\alpha\geq 2$, has one important property such that [11]
\begin{equation}\label{}
g_{\alpha}(\sqrt{x^{2}+y^{2}})\geq g_{\alpha}(x)+g_{\alpha}(y)
\end{equation}
for $0\leq x,y,x^{2}+y^{2}\leq1$.

When $\alpha\geq\frac{\sqrt{7}-1}{2}$, it is easy to see in [12] that the function $g_{\alpha}(x)$ satisfies the following relations
\begin{equation}\label{}
[g_{\alpha}(\sqrt{x^{2}+y^{2}})]^{2}\geq [g_{\alpha}(x)]^{2}+[g_{\alpha}(y)]^{2}
\end{equation}
for $0\leq x,y,x^{2}+y^{2}\leq1$.

\section{TIGHTER MONOGAMY RELATIONS FOR R\'{E}NYI-$\alpha$ ENTANGLEMENT}

In the following, we establish a class of tight R\'{e}nyi-$\alpha$ entanglement monogamy relations related to the power $\eta$. We first provide the following lemma.

\textbf{Lemma 1.} For $x\in [0,1]$ and $\mu\geq1$, then
\begin{equation}\label{}
(1+x)^{\mu}\geq1+\frac{\mu}{2} x+(2^{\mu}-\frac{\mu}{2}-1)x^{\mu}\geq 1+(2^{\mu}-1)x^{\mu}.
\end{equation}

Proof. If $x=0$, the inequality is trivial. Otherwise, let $f(\mu,x)=\frac{(1+x)^{\mu}-\frac{\mu}{2} x-1}{x^{\mu}}$. Then, $\frac{\partial f}{\partial x}=\frac{\mu x^{\mu-1}[1+\frac{(\mu-1)}{2}x-(1+x)^{\mu-1}]}{x^{2\mu}}$. When $\mu\geq1$ and $0\leq x\leq1$, it is obviously that $1+\frac{(\mu-1)}{2}x\leq(1+x)^{\mu-1}$. Thus, $\frac{\partial f}{\partial x}\leq0$, $f(\mu,x)$ is a decreasing function of $x$, i.e. $ f(\mu,x)\geq f(\mu,1)=2^{\mu}-\frac{\mu}{2}-1$. Thus we have $(1+x)^{\mu}\geq1+\frac{\mu}{2} x+(2^{\mu}-\frac{\mu}{2}-1)x^{\mu}$.

Since $x\geq x^{\mu}$, for $x\in [0,1]$ and $\mu\geq1$, one gets $1+\frac{\mu}{2} x+(2^{\mu}-\frac{\mu}{2}-1)x^{\mu}=1+\frac{\mu}{2}(x-x^{\mu})+(2^{\mu}-1)x^{\mu}\geq 1+(2^{\mu}-1)x^{\mu}$. Altogether, we can get $(1+x)^{\mu}\geq1+\frac{\mu}{2} x+(2^{\mu}-\frac{\mu}{2}-1)x^{\mu}\geq 1+(2^{\mu}-1)x^{\mu}$.

Now we provide our main results of this paper.

\textbf{Lemma 2}. For an $N$-qubit state $\rho_{A|B_{1}\cdots B_{N-1}}\in \mathcal{H}_{A}\otimes \mathcal{H}_{B_{1}}\otimes\cdots\otimes \mathcal{H}_{B_{N-1}}$, if $C(\rho_{A|B_{i}})\geq C(\rho_{A|B_{i+1}\cdots B_{N-1}})$ for $i=1,2,\cdots,N-2$, then
\begin{equation}\label{}
 \begin{aligned}
 E_{\alpha}^{\eta}(\rho_{A|B_{1}\cdots B_{N-1}})& \geq
 E_{\alpha}^{\eta}(\rho_{A|B_{1}})+ (2^{\eta}-1)E_{\alpha}^{\eta}(\rho_{A|B_{2}})+\cdots+(2^{\eta}-1)^{N-4}E_{\alpha}^{\eta}(\rho_{A|B_{N-3}})\\
 & \quad +(2^{\eta}-1)^{N-3}\left\{E_{\alpha}^{\eta}(\rho_{A|B_{N-2}})+\frac{\eta}{2}E_{\alpha}^{\eta-1}(\rho_{A|B_{N-2}})E_{\alpha}(\rho_{A|B_{N-1}})
 +(2^{\eta}-\frac{\eta}{2}-1)E_{\alpha}^{\eta}(\rho_{A|B_{N-1}})\right\}
\end{aligned}
\end{equation}
for $\alpha\geq 2$ and the power $\eta\geq1$.

Proof. From the inequality (8), for $\eta\geq1$, we have
\begin{equation}\label{}
[g_{\alpha}(\sqrt{x^{2}+y^{2}})]^{\eta}\geq [g_{\alpha}(x)+g_{\alpha}(y)]^{\eta}.
\end{equation}

Without loss of generality, we assume $x\geq y$, then we obtain
\begin{equation}\label{}
[g_{\alpha}(\sqrt{x^{2}+y^{2}})]^{\eta}\geq [g_{\alpha}(x)]^{\eta}+\frac{\eta}{2} [g_{\alpha}(x)]^{\eta-1}[g_{\alpha}(y)]+(2^{\eta}-\frac{\eta}{2}-1)[g_{\alpha}(y)]^{\eta}.
\end{equation}
Here the inequality is due to Lemma 1.

Let us first consider an $N$-qubit pure state $|\Psi\rangle_{A|B_{1}\cdots B_{N-1}}$. The entanglement $E_{\alpha}(|\Psi\rangle_{A|B_{1}\cdots B_{N-1}})$ and $C(|\Psi\rangle_{A|B_{1}\cdots B_{N-1}})$ are related by the function $g_{\alpha}(x)$ in Eq. (7) since the subsystem $B_{1}\cdots B_{N-1}$ can be regarded as a logic qubit. Thus, we can obtain\begin{equation}\label{}
 \begin{aligned}
  E_{\alpha}^{\eta}(|\Psi\rangle_{A|B_{1}\cdots B_{N-1}}) & =
  [g_{\alpha}(C(|\Psi\rangle_{A|B_{1}\cdots B_{N-1}}))]^{\eta}\\
  & \geq
  \left [g_{\alpha}\left (\sqrt{C^{2}(\rho_{A|B_{1}})+\cdots+C^{2}(\rho_{A|B_{N-1}})} \right )\right ]^{\eta}\\
  & \geq
  [g_{\alpha}(C(\rho_{A|B_{1}}))]^{\eta}+\frac{\eta}{2} [g_{\alpha}(C(\rho_{A|B_{1}}))]^{\eta-1}g_{\alpha}\left(\sqrt{C^{2}(\rho_{A|B_{2}})+\cdots+C^{2}(\rho_{A|B_{N-1}})}\right)\\
   & \quad+(2^{\eta}-\frac{\eta}{2}-1)\left [g_{\alpha}\left(\sqrt{C^{2}(\rho_{A|B_{2}})+\cdots+C^{2}(\rho_{A|B_{N-1}})}\right)\right]^{\eta}\\
   & \geq
 \cdots\\
  & \geq
  [g_{\alpha}(C(\rho_{A|B_{1}}))]^{\eta}+(2^{\eta}-1)[g_{\alpha}(C(\rho_{A|B_{2}}))]^{\eta}
  +\cdots+(2^{\eta}-1)^{N-4}[g_{\alpha}(C(\rho_{A|B_{N-3}}))]^{\eta}\\
  & \quad+(2^{\eta}-1)^{N-3}\left\{[g_{\alpha}(C(\rho_{A|B_{N-2}}))]^{\eta}
  +\frac{\eta}{2}[g_{\alpha}(C(\rho_{A|B_{N-2}}))]^{\eta-1}g_{\alpha}(C(\rho_{A|B_{N-1}}))\right.\\
  & \quad+\left.(2^{\eta}-\frac{\eta}{2}-1)[g_{\alpha}(C(\rho_{A|B_{N-1}}))]^{\eta}\right\}\\
  & =
   E_{\alpha}^{\eta}(\rho_{A|B_{1}})+ (2^{\eta}-1)E_{\alpha}^{\eta}(\rho_{A|B_{2}})+\cdots+(2^{\eta}-1)^{N-4}E_{\alpha}^{\eta}(\rho_{A|B_{N-3}})
   \\
  & \quad +(2^{\eta}-1)^{N-3}\left\{E_{\alpha}^{\eta}(\rho_{A|B_{N-2}})+\frac{\eta}{2}E_{\alpha}^{\eta-1}(\rho_{A|B_{N-2}})E_{\alpha}(\rho_{A|B_{N-1}})\right.\\
  & \quad +\left.(2^{\eta}-\frac{\eta}{2}-1)E_{\alpha}^{\eta}(\rho_{A|B_{N-1}})\right\},
\end{aligned}
\end{equation}
where we have utilized the monogamy inequality (5) and the monotonically increasing property of the function $g_{\alpha}(x)$ to obtain the the first inequality, the second inequality is due to Eq. (13) by letting $x=C(\rho_{A|B_{1}})$ and $y=\sqrt{C^{2}(\rho_{A|B_{2}})+\cdots+C^{2}(\rho_{A|B_{N-1}})}$, the other inequalities are from iterative use of Eq. (13) and  Eq. (8). In fact, we also use the conditions $1+\frac{\mu}{2} x+(2^{\mu}-\frac{\mu}{2}-1)x^{\mu}\geq 1+(2^{\mu}-1)x^{\mu}$ and $C(\rho_{A|B_{i}})\geq C(\rho_{A|B_{i+1}\cdots B_{N-1}})\geq \sqrt{C^{2}(\rho_{A|B_{i+1}})+\cdots+C^{2}(\rho_{A|B_{N-1}})}$, $i=1,2,\cdots,N-2$. Since for any two-qubit state $\rho_{AB}$, when $\alpha\geq\frac{\sqrt{7}-1}{2}$, $E_{\alpha}(\rho_{AB})=g_{\alpha}(C(\rho_{AB}))$, we obtain the last equality.

Next, let us consider an $N$-qubit mixed state $\rho_{A|B_{1}\cdots B_{N-1}}$. Suppose that the optimal decomposition for $E_{\alpha}(\rho_{A|B_{1}\cdots B_{N-1}})$ is $\rho_{A|B_{1}\cdots B_{N-1}}=\sum_{k}p_{k}|\varphi_{k}\rangle_{A|B_{1}\cdots B_{N-1}}\langle\varphi_{k}|\in \mathcal{H}_{A}\otimes \mathcal{H}_{B_{1}}\otimes\cdots\otimes \mathcal{H}_{B_{N-1}}$. Thus, we have
\begin{equation}\label{}
 \begin{aligned}
  E_{\alpha}(\rho_{A|B_{1}\cdots B_{N-1}}) & = \sum_{k}p_{k}E_{\alpha}(|\varphi_{k}\rangle_{A|B_{1}\cdots B_{N-1}})\\
  & = \sum_{k}p_{k}g_{\alpha}(C(|\varphi_{k}\rangle_{A|B_{1}\cdots B_{N-1}}))\\
  & \geq
  g_{\alpha}\left (\sum_{k}p_{k}C(|\varphi_{k}\rangle_{A|B_{1}\cdots B_{N-1}})\right )\\
  & \geq
  g_{\alpha}\left (\sum_{l}p_{l}C(|\chi_{l}\rangle_{A|B_{1}\cdots B_{N-1}})\right)\\
  & =
  g_{\alpha}(C(\rho_{A|B_{1}\cdots B_{N-1}})),
\end{aligned}
\end{equation}
where the first inequality follows from the convex property of the function $g_{\alpha}(x)$, the second equality is satisfied due to $\{p_{l}, |\chi_{l}\rangle_{A|B_{1}\cdots B_{N-1}}\}$ being the optimal decomposition for $C(\rho_{A|B_{1}\cdots B_{N-1}})$.

Now we can derive
\begin{equation}\label{}
 \begin{aligned}
  E_{\alpha}^{\eta}(\rho_{A|B_{1}\cdots B_{N-1}}) & \geq
  \left[g_{\alpha}(C(\rho_{A|B_{1}\cdots B_{N-1}}))\right]^{\eta}\\
  & \geq
   \left[g_{\alpha}\left (\sqrt{C^{2}(\rho_{A|B_{1}})+\cdots+C^{2}(\rho_{A|B_{N-1}})} \right )\right]^{\eta}\\
  & \geq
  [g_{\alpha}(C(\rho_{A|B_{1}}))]^{\eta}+(2^{\eta}-1)[g_{\alpha}(C(\rho_{A|B_{2}}))]^{\eta}
  +\cdots+(2^{\eta}-1)^{N-4}[g_{\alpha}(C(\rho_{A|B_{N-3}}))]^{\eta}\\
  & \quad+(2^{\eta}-1)^{N-3}\left\{[g_{\alpha}(C(\rho_{A|B_{N-2}}))]^{\eta}
  +\frac{\eta}{2}[g_{\alpha}(C(\rho_{A|B_{N-2}}))]^{\eta-1}g_{\alpha}(C(\rho_{A|B_{N-1}}))\right.\\
  & \quad+\left.(2^{\eta}-\frac{\eta}{2}-1)[g_{\alpha}(C(\rho_{A|B_{N-1}}))]^{\eta}\right\}\\
  & =
   E_{\alpha}^{\eta}(\rho_{A|B_{1}})+ (2^{\eta}-1)E_{\alpha}^{\eta}(\rho_{A|B_{2}})+\cdots+(2^{\eta}-1)^{N-4}E_{\alpha}^{\eta}(\rho_{A|B_{N-3}})
   \\
  & \quad +(2^{\eta}-1)^{N-3}\left\{E_{\alpha}^{\eta}(\rho_{A|B_{N-2}})+\frac{\eta}{2}E_{\alpha}^{\eta-1}(\rho_{A|B_{N-2}})E_{\alpha}(\rho_{A|B_{N-1}})\right.\\
  & \quad+\left.(2^{\eta}-\frac{\eta}{2}-1)E_{\alpha}^{\eta}(\rho_{A|B_{N-1}})\right\}.
\end{aligned}
\end{equation}
Here in the second inequality we have used the monogamy inequality (5) and the monotonically increasing property of the function $g_{\alpha}(x)$. By iterative use of inequality (13), we have the third inequality. As a matter of fact, the conditions $1+\frac{\mu}{2} x+(2^{\mu}-\frac{\mu}{2}-1)x^{\mu}\geq 1+(2^{\mu}-1)x^{\mu}$ and $C(\rho_{A|B_{i}})\geq C(\rho_{A|B_{i+1}\cdots B_{N-1}})\geq \sqrt{C^{2}(\rho_{A|B_{i+1}})+\cdots+C^{2}(\rho_{A|B_{N-1}})}$, $i=1,2,\cdots,N-2$, have been used. Since for any two-qubit state $\rho_{AB}$, when $\alpha\geq\frac{\sqrt{7}-1}{2}$, $E_{\alpha}(\rho_{AB})=g_{\alpha}(C(\rho_{AB}))$, one gets the last equality, and the proof is completed.

Based on the Lemma 2,  if some $C(\rho_{A|B_{i}})\geq C(\rho_{A|B_{i+1}\cdots B_{N-1}})$ and some $C(\rho_{A|B_{j}})\leq C(\rho_{A|B_{j+1}\cdots B_{N-1}})$, we have the following conclusion.

\textbf{Lemma 3.} For an $N$-qubit state $\rho_{A|B_{1}\cdots B_{N-1}}\in \mathcal{H}_{A}\otimes \mathcal{H}_{B_{1}}\otimes\cdots\otimes \mathcal{H}_{B_{N-1}}$, if $C(\rho_{A|B_{i}})\geq C(\rho_{A|B_{i+1}\cdots B_{N-1}})$ for $i=1,2,\cdots m$, and $C(\rho_{A|B_{j}})\leq C(\rho_{A|B_{j+1}\cdots B_{N-1}})$ for $j=m+1,\cdots,N-2,\forall\ 1\leq m\leq N-3,N\geq4$, then
\begin{equation}\label{}
 \begin{aligned}
 E_{\alpha}^{\eta}(\rho_{A|B_{1}\cdots B_{N-1}}) \geq &
 E_{\alpha}^{\eta}(\rho_{A|B_{1}})+(2^{\eta}-1)E_{\alpha}^{\eta}(\rho_{A|B_{2}})+\cdots
 +(2^{\eta}-1)^{m-1}E_{\alpha}^{\eta}(\rho_{A|B_{m}})\\
& +(2^{\eta}-1)^{m+1}[E_{\alpha}^{\eta}(\rho_{A|B_{m+1}})+\cdots+E_{\alpha}^{\eta}(\rho_{A|B_{N-3}})]\\
& +(2^{\eta}-1)^{m}\left\{(2^{\eta}-\frac{\eta}{2}-1)[E_{\alpha}(\rho_{A|B_{N-2}})]^{\eta}+
\frac{\eta}{2}E_{\alpha}(\rho_{A|B_{N-2}})E_{\alpha}^{\eta-1}(\rho_{A|B_{N-1}})\right.\\
& +\left.[E_{\alpha}(\rho_{A|B_{N-1}})]^{\eta}\right\}
\end{aligned}
\end{equation}
for $\alpha\geq2$ and the power $\eta\geq1$.

Proof. From Lemma 2, we have
\begin{equation}\label{}
 \begin{aligned}
  E_{\alpha}^{\eta}(\rho_{A|B_{1}\cdots B_{N-1}}) & \geq
  [g_{\alpha}(C(\rho_{A|B_{1}}))]^{\eta}+(2^{\eta}-1)[g_{\alpha}(C(\rho_{A|B_{2}}))]^{\eta}+\cdots+(2^{\eta}-1)^{m-2}[g_{\alpha}(C(\rho_{A|B_{m-1}}))]^{\eta}\\
  & \quad+(2^{\eta}-1)^{m-1}\left\{[g_{\alpha}(C(\rho_{A|B_{m}}))]^{\eta}+\frac{\eta}{2}[g_{\alpha}(C(\rho_{A|B_{m}}))]^{\eta-1}g_{\alpha}(C(\rho_{A|B_{m+1}\cdots B_{N-1}}))\right.\\
  & \quad \left.+(2^{\eta}-\frac{\eta}{2}-1)[g_{\alpha}(C(\rho_{A|B_{m+1}\cdots B_{N-1}}))]^{\eta}\right\}\\
  & \geq
  [g_{\alpha}(C(\rho_{A|B_{1}}))]^{\eta}+(2^{\eta}-1)[g_{\alpha}(C(\rho_{A|B_{2}}))]^{\eta}+\cdots+(2^{\eta}-1)^{m-1}[g_{\alpha}(C(\rho_{A|B_{m}}))]^{\eta}\\
  & \quad+(2^{\eta}-1)^{m}[g_{\alpha}(C(\rho_{A|B_{m+1}\cdots B_{N-1}}))]^{\eta}.
  \end{aligned}
\end{equation}
With a similar procedure as $C(\rho_{A|B_{j}})\leq C(\rho_{A|B_{j+1}\cdots B_{N-1}})$ for $j=m+1,\cdots,N-2$, we have
\begin{equation}\label{}
 \begin{aligned}
\left[g_{\alpha}(C(\rho_{A|B_{m+1}\cdots B_{N-1}}))\right]^{\eta} & \geq
(2^{\eta}-\frac{\eta}{2}-1)[g_{\alpha}(C(\rho_{A|B_{m+1}}))]^{\eta}+\frac{\eta}{2}g_{\alpha}(C(\rho_{A|B_{m+1}}))[g_{\alpha}(C(\rho_{A|B_{m+2}\cdots B_{N-1}}))]^{\eta-1}\\
& \quad+[g_{\alpha}(C(\rho_{A|B_{m+2}\cdots B_{N-1}}))]^{\eta}\\
& \geq
(2^{\eta}-1)\left\{[g_{\alpha}(C(\rho_{A|B_{m+1}}))]^{\eta}+\cdots+[g_{\alpha}(C(\rho_{A|B_{N-3}}))]^{\eta}\right\}\\
& \quad+(2^{\eta}-\frac{\eta}{2}-1)[g_{\alpha}(C(\rho_{A|B_{N-2}}))]^{\eta}+\frac{\eta}{2}g_{\alpha}(C(\rho_{A|B_{N-2}}))[g_{\alpha}(C(\rho_{A|B_{N-1}}))]^{\eta-1}\\
& \quad+[g_{\alpha}(C(\rho_{A|B_{N-1}}))]^{\eta}.
\end{aligned}
\end{equation}
Combining inequalities (18) and (19), we have Lemma 3.

Now, we present a tight monogamy relation for $2>\alpha\geq\frac{\sqrt{7}-1}{2}$.

\textbf{Lemma 4}.  For an $N$-qubit state $\rho_{A|B_{1}\cdots B_{N-1}}\in \mathcal{H}_{A}\otimes \mathcal{H}_{B_{1}}\otimes\cdots\otimes \mathcal{H}_{B_{N-1}}$, if $C(\rho_{A|B_{i}})\geq C(\rho_{A|B_{i+1}\cdots B_{N-1}})$ for $i=1,2,\cdots,N-2$, then
\begin{equation}\label{}
 \begin{aligned}
  E_{\alpha}^{\eta}(\rho_{A|B_{1}\cdots B_{N-1}}) \geq & E_{\alpha}^{\eta}(\rho_{A|B_{1}})+(2^{t}-1)E_{\alpha}^{\eta}(\rho_{A|B_{2}})+\cdots+(2^{t}-1)^{N-4}E_{\alpha}^{\eta}(\rho_{A|B_{N-3}})\\
 & +(2^{t}-1)^{N-3}\left\{E_{\alpha}^{\eta}(\rho_{A|B_{N-2}})+\frac{t}{2}E_{\alpha}^{\eta-2}(\rho_{A|B_{N-1}})E_{\alpha}^{2}(\rho_{A|B_{N-2}})\right.\\
 & +\left.(2^{t}-\frac{t}{2}-1)E_{\alpha}^{\eta}(\rho_{A|B_{N-1}})\right\}
 \end{aligned}
\end{equation}
for $2>\alpha\geq\frac{\sqrt{7}-1}{2}$ and the power $\eta\geq 2$, where $t=\frac{\eta}{2}$.

Proof. From the inequality (9), for $\eta\geq 2$, we have
\begin{equation}\label{}
[g_{\alpha}(\sqrt{x^{2}+y^{2}})]^{\eta}\geq \left\{[g_{\alpha}(x)]^{2}+[g_{\alpha}(y)]^{2}\right\}^{t}.
\end{equation}

Without loss of generality, we assume $x\geq y$, according to Lemma 1, we have
\begin{equation}\label{}
[g_{\alpha}(\sqrt{x^{2}+y^{2}})]^{\eta}\geq [g_{\alpha}(x)]^{\eta}+\frac{t}{2}[g_{\alpha}(x)]^{\eta-2}[g_{\alpha}(y)]^{2}+(2^{t}-\frac{t}{2}-1)[g_{\alpha}(y)]^{\eta}.
\end{equation}

By using inequalities (13) and (14) and the similar consideration in the proof of Lemma 2, we obtain Lemma 4. If some $C(\rho_{A|B_{i}})\geq C(\rho_{A|B_{i+1}\cdots B_{N-1}})$ and some $C(\rho_{A|B_{j}})\leq C(\rho_{A|B_{j+1}\cdots B_{N-1}})$, then we have the following lemma.

\textbf{Lemma 5}. For an $N$-qubit state $\rho_{A|B_{1}\cdots B_{N-1}}\in \mathcal{H}_{A}\otimes \mathcal{H}_{B_1}\otimes\cdots\otimes \mathcal{H}_{B_{N-1}}$, if $C(\rho_{A|B_{i}})\geq C(\rho_{A|B_{i+1}\cdots B_{N-1}})$ for $i=1,2,\cdots,m$, and $C(\rho_{A|B_{j}})\leq C(\rho_{A|B_{j+1}\cdots B_{N-1}})$ for $j=m+1,\cdots,N-2,\forall\ 1\leq m\leq N-3,N\geq4$, then
\begin{equation}\label{}
 \begin{aligned}
  E_{\alpha}^{\eta}(\rho_{A|B_{1}\cdots B_{N-1}}) \geq & E_{\alpha}^{\eta}(\rho_{A|B_{1}})+(2^{t}-1)E_{\alpha}^{\eta}(\rho_{A|B_{2}})+\cdots
  +(2^{t}-1)^{m-1}E_{\alpha}^{\eta}(\rho_{A|B_{m}})\\
  &+(2^{t}-1)^{m+1}[E_{\alpha}^{\eta}(\rho_{A|B_{m+1}})+\cdots+E_{\alpha}^{\eta}(\rho_{A|B_{N-3}})]\\
  &+(2^{t}-1)^{m}\left\{(2^{t}-\frac{t}{2}-1)E_{\alpha}^{\eta}(\rho_{A|B_{N-2}})+
  \frac{t}{2}E_{\alpha}^{2}(\rho_{A|B_{N-2}})E_{\alpha}^{\eta-2}(\rho_{A|B_{N-1}})\right.\\
  &+\left.E_{\alpha}^{\eta}(\rho_{A|B_{N-1}})\right\}
 \end{aligned}
\end{equation}
for $2>\alpha\geq\frac{\sqrt{7}-1}{2}$ and the power $\eta\geq 2$, where $t=\frac{\eta}{2}$.

We note that for $\eta> 2$, $t=\frac{\eta}{2}> 1$, $1+\frac{t}{2} x+(2^t-\frac{t}{2}-1)x^t> 1+(2^t-1)x^t$ for all $x\in (0,1)$, Lemma 4 and Lemma 5 give a tighter monogamy relation with larger lower bounds than the result in [12, 15, 16].

\section{CONCLUSION}

As a fundamental problem in quantum entanglement theory, multipartite entanglement has attracted increasing interest over the last 20 years. We have investigated the monogamy relations related to the power of the R\'{e}nyi-$\alpha$ entanglement for $N$-qubit state. When $\alpha\geq2$ and the power $\eta>1$, we derive a tighter monogamy relation than the monogamy relations just based on the power of the R\'{e}nyi-$\alpha$ entanglement. For $2>\alpha\geq\frac{\sqrt{7}-1}{2}$ and the power $\eta>2$, we give a tighter monogamy relation with larger lower bounds than the result in [12, 15, 16]. Our result can provide a useful methodology to study further the monogamy properties of the multiparty quantum entanglement.

\vspace{0.6cm}
\acknowledgments
This work was supported by the National Natural Science Foundation of China under Grant No: 11475054, the Hebei Natural Science Foundation of China under Grant No: A2018205125.

\end{document}